\documentclass[runningheads]{llncs}

\usepackage{graphicx}
\usepackage{subcaption}
\usepackage[utf8]{inputenc}

\begin{document}
\title{Automatic Polyp Segmentation Using Convolutional Neural Networks}
%
%
\author{Sara Hosseinzadeh Kassani\inst{1} \orcidID{0000-0002-5776-7929} \and
Peyman Hosseinzadeh Kassani\inst{2} \orcidID{0000-0001-7736-3959} \and
Michal J. Wesolowski\inst{3} \and
Kevin A. Schneider\inst{1} \and
Ralph Deters\inst{1}}
\authorrunning{S. Hosseinzadeh Kassani et al.}
%
\institute{University of Saskatchewan, Department of Computer Science, Saskatoon, Canada \and
University of Stanford, Department of Neurology and Neurological, California, USA \and
University of Saskatchewan, Department of Medical Imaging, Saskatoon, Canada}

\maketitle              
\begin{abstract}
Colorectal cancer is the third most common cancer-related death after lung cancer and breast cancer worldwide. The risk of developing colorectal cancer could be reduced by early diagnosis of polyps during a colonoscopy. Computer-aided diagnosis systems have the potential to be applied for polyp screening and reduce the number of missing polyps. In this paper, we compare the performance of different deep learning architectures as feature extractors, i.e. ResNet, DenseNet, InceptionV3, InceptionResNetV2 and SE-ResNeXt in the encoder part of a U-Net architecture. We validated the performance of presented ensemble models on the CVC-Clinic (GIANA 2018) dataset. The DenseNet169 feature extractor combined with U-Net architecture outperformed the other counterparts and achieved an accuracy of 99.15\%, Dice similarity coefficient of 90.87\%, and Jaccard index of 83.82\%. 

\end{abstract}
\section{Introduction}

\begin{figure}[ht]
	\centering
	\begin{subfigure}{0.2\textwidth}
		\includegraphics[width=\linewidth]{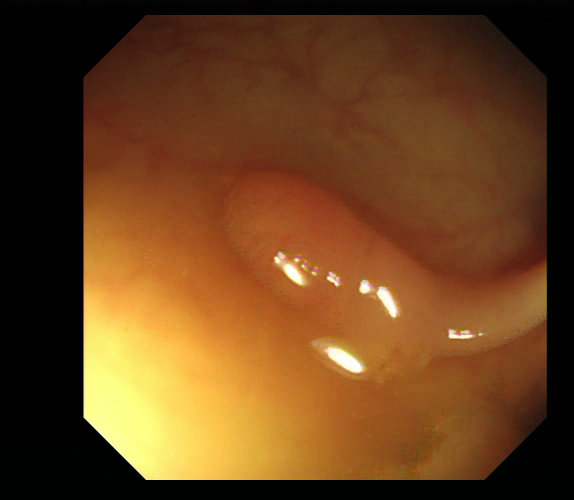}
	\end{subfigure}\hfil
	\begin{subfigure}{0.2\textwidth}
		\includegraphics[width=\linewidth]{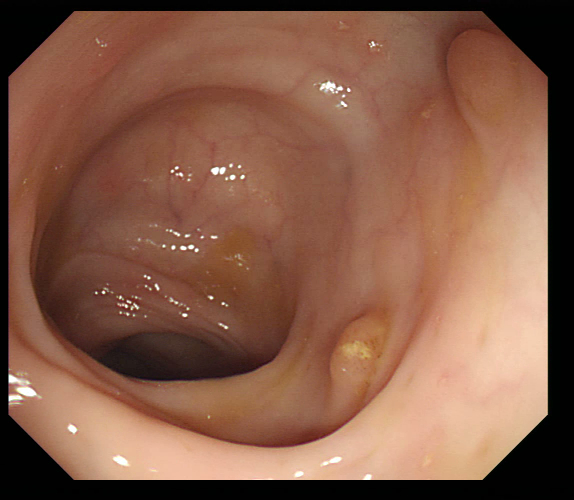}
	\end{subfigure}\hfil
	\begin{subfigure}{0.2\textwidth}
		\includegraphics[width=\linewidth]{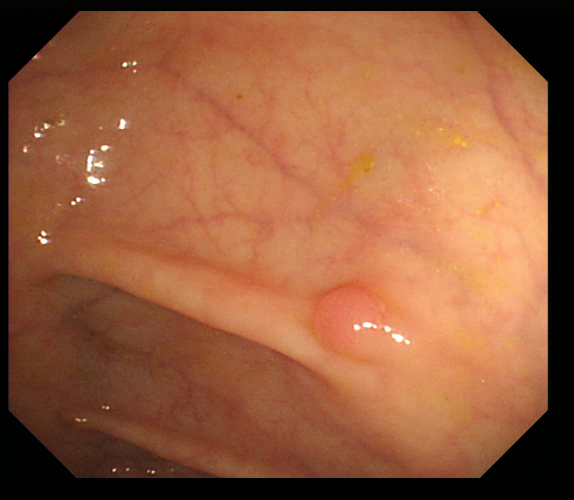}
	\end{subfigure}\hfil	
	
	\medskip
	\begin{subfigure}{0.2\textwidth}
		\includegraphics[width=\linewidth]{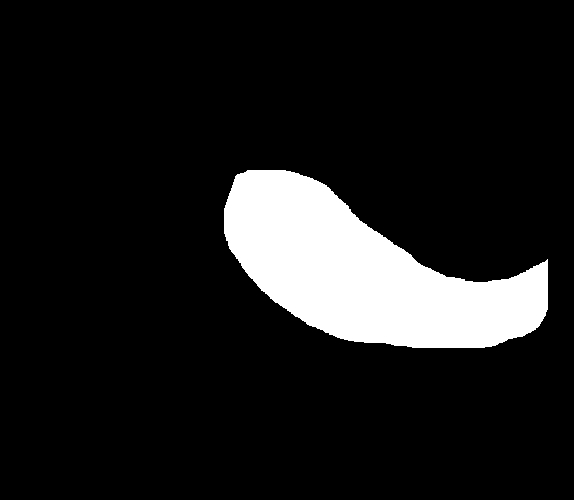}
	\end{subfigure}\hfil
	\begin{subfigure}{0.2\textwidth}
		\includegraphics[width=\linewidth]{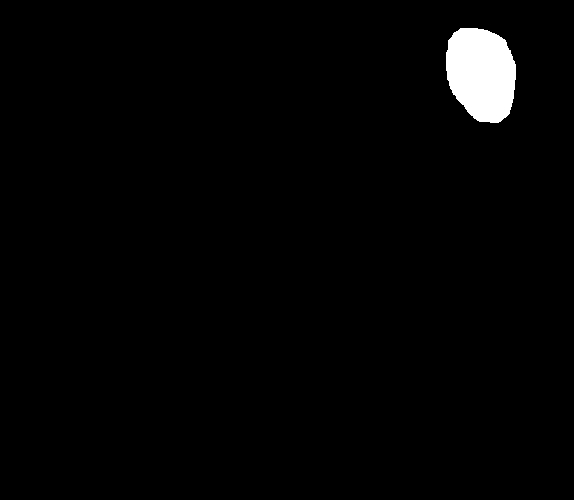}
	\end{subfigure}\hfil
	\begin{subfigure}{0.2\textwidth}
		\includegraphics[width=\linewidth]{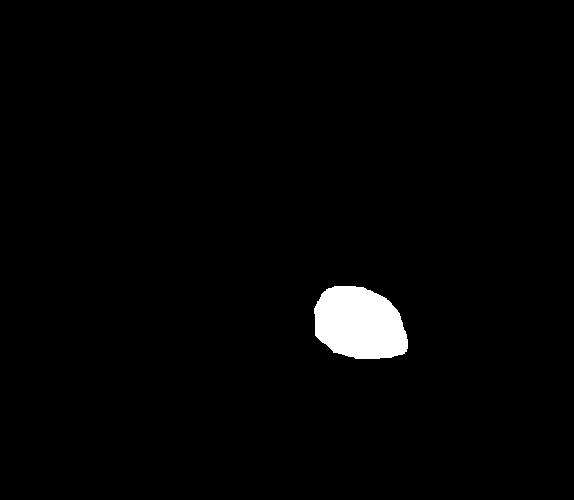}
	\end{subfigure}\hfil
	
	\caption{Some examples of polyps from colonoscopy images (first row) and their corresponding manual segmentations provided by expert endoscopists (second row).}
	\label{fig:SegmentedDataExamples}
\end{figure}

Colorectal cancer is the third most common cancer-related death in the United States in both men and women. According to the annual report provided by American cancer society~\cite{ColorectalCancerStatistics}, approximately 101,420 new cases of colon cancer and 44,180 new cases of rectal cancer will be diagnosed in 2019. Additionally, 51,020 patients are expected to die from colorectal cancer during 2019 in the United States. Most colorectal cancers start  as benign polyps in the inner linings of the colon or rectum. Removal of these polyps can decrease the risk of developing cancer. Colonoscopy is the gold standard for screening and detecting polyps~\cite{akbari2018classification}. Screening and analysis  of polyps in colonoscopy images is dependent on experienced endoscopists~\cite{qadir2019polyp}. Polyp detection is considered as a challenging task due to the variations in size and shape of polyps among different patients. This is illustrated in Fig.~\ref{fig:SegmentedDataExamples}, where the segmented regions vary in size, shape and position. 

\begin{figure}[h]
	\centering
	\includegraphics[width=0.58\linewidth]{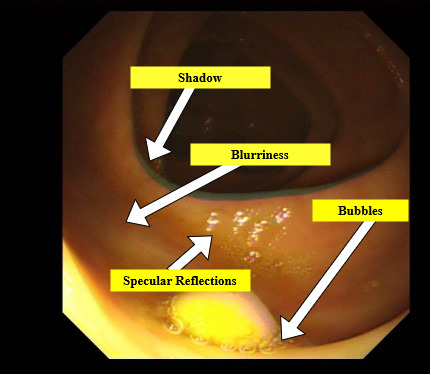}
	\caption{Examples of different noises exist in colonoscopy images.}
	\label{fig:occlusionpolyp}
\end{figure}

The miss rates of smaller polyps during the colonoscopy is also another issue that needs to be addressed. Developing computer-aided diagnosis (CAD) systems can assist physicians in the early detection of polyps. CAD systems using convolutional neural networks (CNN) is an active research area and has the potential to reduce polyp miss rate~\cite{qadir2019improving}. Recent developments based on the application of deep learning-based techniques achieved promising results for the segmentation and extraction of polyps and improved the detection rate, despite the complexity of the case during colonoscopy~\cite{nguyen2018colorectal}~\cite{huang2018automatic}~\cite{wickstrom2018uncertainty}~\cite{li2017colorectal}. The presence of visual occlusions such as shadows, reflections, blurriness and illumination conditions, as shown in Fig.~\ref{fig:occlusionpolyp} can adversely affect the performance of CNN and the quality of the segmented polyp region. 

\subsection{Motivation and contributions}

The main motivation of this paper is to compare the performance of different CNN modules, i.e. Squeeze-and-Excitation (SE) blocks, inception blocks, residual blocks and dense blocks for building automatic polyp segmentation systems. Considering the problem of intra- and inter-observer variability, designing hand-crafted features with limited representation capability requires expert knowledge and extensive application-specific fine-tuning. Also, employing very deep networks for small data samples suffers from gradient vanishing and poor local minima issues. In this study, we evaluate the performance of different CNN architectures (e.g., ResNet~\cite{he2016deep}, DenseNet~\cite{huang2017densely}, InceptionV3~\cite{szegedy2016rethinking}, InceptionResNet~\cite{szegedy2017inceptionV4}, SE-ResNeXt~\cite{hu2018squeeze}) with various modules as feature extractor to the encoder part of a U-Net architecture to investigate the impact of incorporating modules in extracting high-level contextual information from the input image. In this way, we provide better insights on how different convolutional pathways efficiently incorporate both local and contextual image information for training producers and cope with the inherent variability of medical data. We validated the performance of presented ensemble models using the CVC-ClinicDB (GIANA 2018) dataset. 

The rest of this study is organized as follows. Section 2 provides the related work in the literature on polyp segmentation approaches. Section 3 presents a detailed description of materials and the methodology. Section 4 describes experimental analysis and discussion of the performance of the segmentation models. Finally, Section 5 concludes the paper and provides future directions.

\section{Related Work} 
Li et al.~\cite{li2017colorectal} presented a fully convolutional neural network for polyp segmentation. The feature extraction stage consists of 8 convolution layers, and 5 pooling layers. The presented method evaluated on CVC-ClinicDB. Li et al approach obtained an accuracy of 96.98\%, f1-score of 83.01\%, sensitivity of 77.32\% and specificity of 99.05\%. 

Akbari et al.~\cite{akbari2018polyp} applied a fully convolutional neural network (FCN-8S) for polyp segmentation. An image patch selection method used for training procedure. Also, a post-processing method (Otsu thresholding) employed on the probability map to improve the performance of the proposed method on the CVC-ColonDB~\cite{CVC-ColonDBLink} database. Akbari et al method achieved an accuracy of 97.70\% and a Dice score of 81.00\%. 


Qadir et al.~\cite{qadir2019polyp} trained a Mask R-CNN with different CNN architectures (Resnet50, Resnet101 and InceptionResnetV2) as a feature extractor for polyp detection and segmentation. Also, two ensemble models of (ensemble of Resnet50 and Resnet101) and (ensemble of Resnet50 and InceptionResnetV2) were employed on CVC-ColonDB dataset. Qadir’s approach achieved 72.59\% recall, 80.00\% precision, 70.42\% Dice score, and 61.24\% Jaccard index.

Nguyen and Lee~\cite{nguyen2018colorectal} used a deep encoder-decoder network method for polyp segmentation from colonoscopy images. The presented encoder-decoder structure consists of atrous convolution and depthwise separable convolution. To improve the performance, the proposed model pre-trained with the VOC 2012 dataset and achieved 88.9\% of Dice score and 89.35\% of Jaccard index on the CVC-ColonDB database.

Kang and Gwak~\cite{kang2019ensemble} employed Mask R-CNN to segment polyp regions in colonoscopy images. Also, an ensemble Mask R-CNN model with different backbone structures (ResNet50 and ResNet101) was adopted to further improve the model performance. The Mask R-CNN was first trained on the COCO dataset and then fine-tuned for polyp segmentation. Three datasets, i.e. CVC-ClinicDB, ETIS-Larib, and CVC-ColonDB, used to measure the performance of the proposed model. The best result achieved on the CVC-ColonDB dataset with 77.92\% mean pixel precision, 76.25\% mean pixel recall and 69.4\% intersection over the union.

\section{Methods and Materials}
\subsection{Experimental Dataset}
In this paper, CVC-ClinicDB~\cite{vazquez2017benchmark}~\cite{bernal2012towards}~\cite{bernal2015wm} database, publicly available at~\cite{GianaDataset}, is used to validate the performance of the presented method. The database consists of 300 Standard Definition (SD) colonoscopy images with a resolution of $574\times500$ pixels, and each image contains one polyp. Each frame has a corresponding ground truth of the region covered by the polyp.

\subsection{Data Pre-processing}
\subsubsection{Resizing:} Regarding to the black margin of each image as illustrated in Fig 3, we center-cropped all images of SD-CVC-ClinicDB from the original size of $574\times500$ pixels to the appropriate size $500\times500$ pixels using bicubic interpolation to reduce the non-informative adjacent background regions.

\subsubsection{Data augmentation:} 
\begin{figure}[t]
	\centering
	\begin{subfigure}{0.28\textwidth}
		\includegraphics[width=\linewidth]{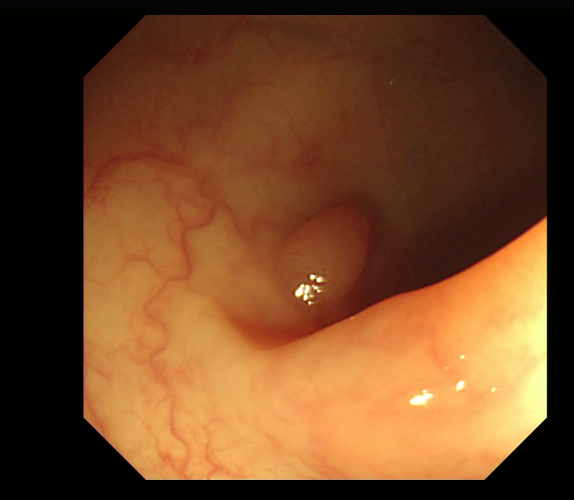}
		\caption{Original image}
		\label{fig:original}
	\end{subfigure}\hfil
	\begin{subfigure}{0.28\textwidth}
		\includegraphics[width=\linewidth]{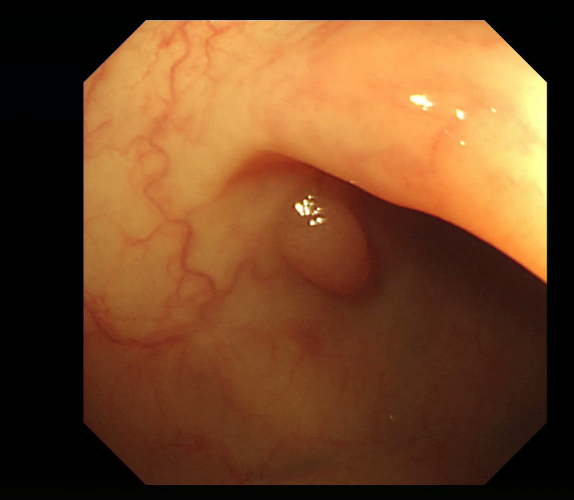}
		\caption{Vertical Flip}
		\label{fig:VF}
	\end{subfigure}\hfil
	\begin{subfigure}{0.28\textwidth}
		\includegraphics[width=\linewidth]{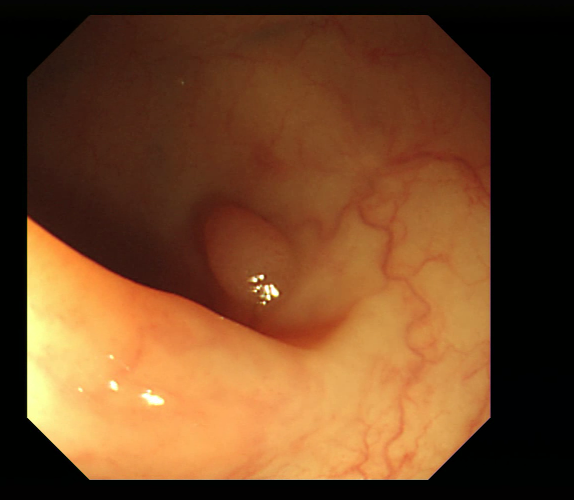}
		\caption{Horizontal Flip}
		\label{fig:HF}
	\end{subfigure}\hfil	
	
	\medskip
	\begin{subfigure}{0.28\textwidth}
		\includegraphics[width=\linewidth]{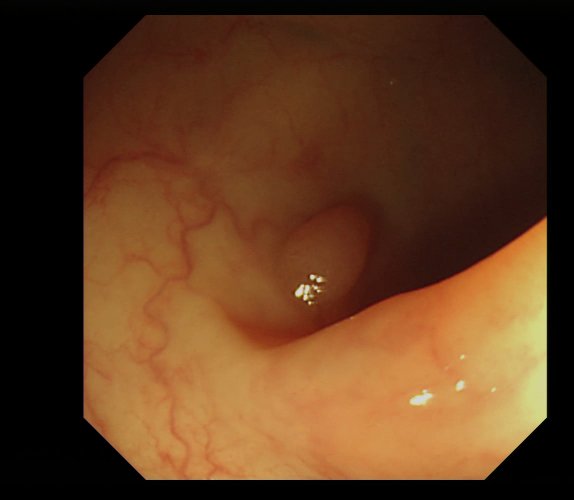}
		\caption{Random Filter}
		\label{fig:RF}
	\end{subfigure}\hfil
	\begin{subfigure}{0.28\textwidth}
		\includegraphics[width=\linewidth]{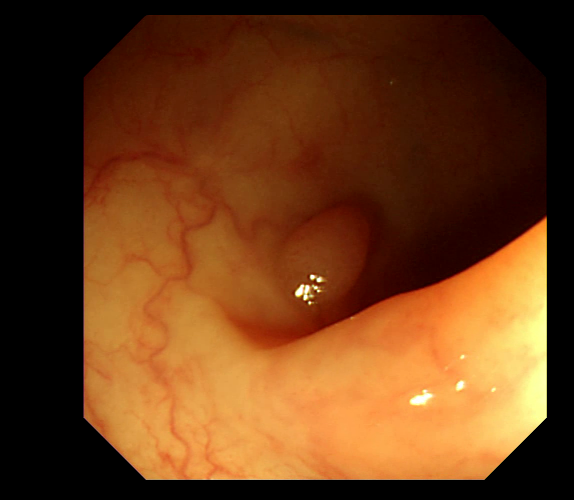}
		\caption{Random Contrast}
		\label{fig:RC}
	\end{subfigure}\hfil
	\begin{subfigure}{0.28\textwidth}
		\includegraphics[width=\linewidth]{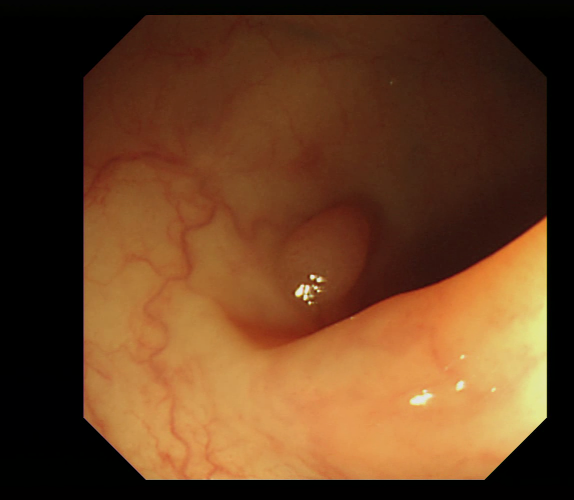}
		\caption{Random Brightness}
		\label{fig:RB}
	\end{subfigure}\hfil
	
	\caption{Examples of data augmentation methods.}
	\label{fig:dataExamples}
\end{figure}

Recent works have demonstrated the advantages of data augmentation methods in extending the size of training data to cover all of the data variances. In this regard, various data augmentation techniques such as horizontal and vertical flipping, rotating and zooming are applied to enlarge the dataset and aid to successfully accomplish segmentation task. Fig.~\ref{fig:dataExamples} shows the examples of the original polyp image (Fig.~\ref{fig:original}) after applying different data augmentation methods. The used methods of augmentation are vertical flipping (Fig 3.b), horizontal flipping (Fig 3.c), random filter such as blur, sharpen (Fig 3.d), random contrast by a factor of 0.5 (Fig 3.e), and finally, random brightness by a factor of 0.5 (Fig 3.f). 

\subsubsection{Z-score Normalization:} To have a uniform distribution from input images and remove bias from input features, we re-scaled the intensity values of the input images to have a zero mean and a standard deviation of one to standardize the input images. 
\subsubsection{Image Normalization:} Before feeding images into the CNN models, we also normalize the intensity values of input images using ImageNet mean subtraction~\cite{krizhevsky2012imagenet}. The ImageNet mean is a pre-computed constant derived from ImageNet~\cite{deng2009imagenet} database.

\subsection{Feature extraction Using Transfer Learning Strategy}
The intuition behind transfer learning is that knowledge learned by a cross-domain dataset transfer into the new dataset in another domain. The main advantages of transfer learning are the improvement of the network performance, reducing the issue of over-fitting, reducing the computational cost, and also the acceleration of the convergence of the network~\cite{lu2019pathological}. In this approach, instead of training a model from scratch, the weights trained on ImageNet dataset or other similar cross-domain dataset is used to initialize weights for the current task. Providing training data large enough to sufficiently train a CNN model is limited due to privacy concerns, which is a common issue in the medical domain. To address the issue of insufficient training samples, transfer learning strategy has also been widely used for accurate and automatic feature extraction in developing various CAD systems. 
\subsection{Ensemble Method}
\subsubsection{U-Net Architecture}
U-Net, proposed by Ronneberger et al.~\cite{ronneberger2015u} in 2015, is an encoder-decoder convolutional network that won ISBI cell tracking challenge. The encoder or down-sampling layers of U-Net architecture learn the feature maps and the decoder or up-sampling layers provide precise segmentation. The encoder part has alternating convolutional filters and max-pooling layers with ReLU activation function to down-sample the data. When the input image is fed into the network, representative features are produced by convolutions at each layer. 
\begin{figure}[hb]
	\centering
	\includegraphics[width=0.98\linewidth]{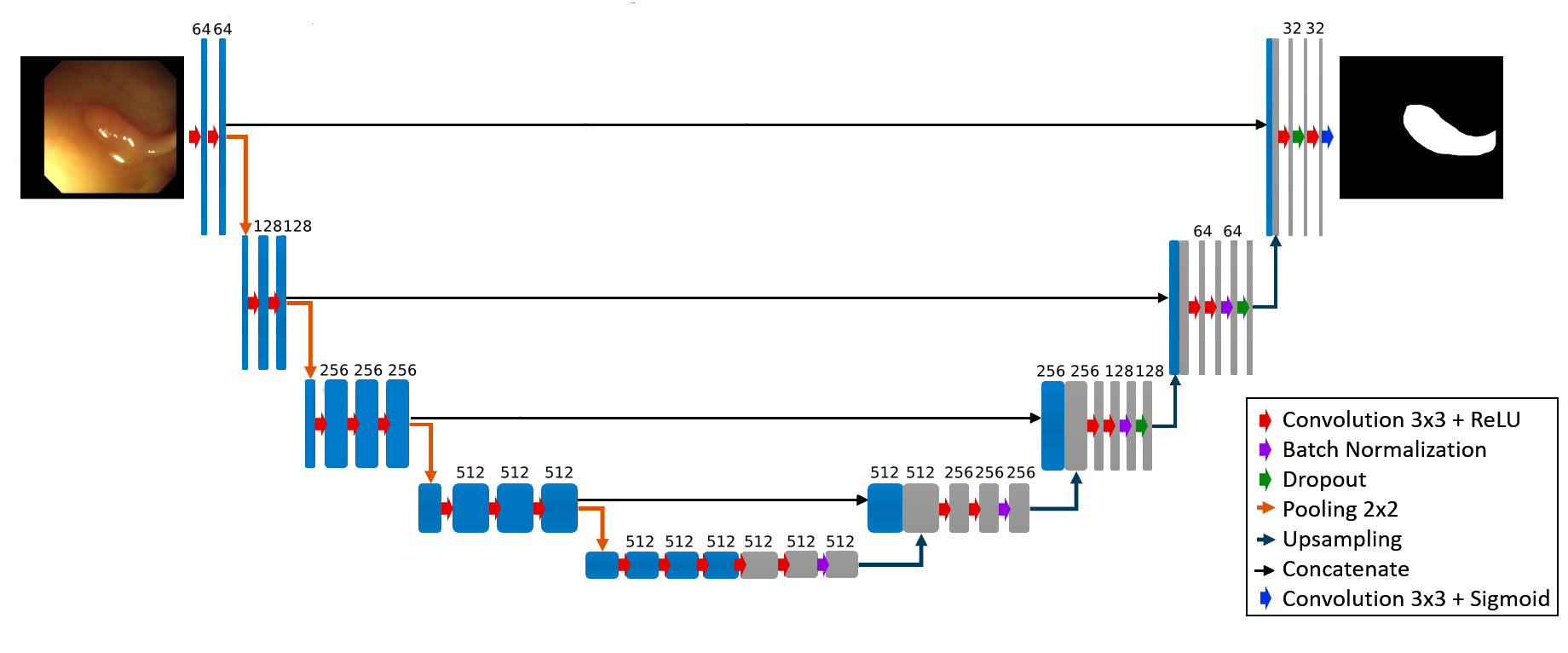}
	\caption{Proposed Approach for polyp segmentation. The CNN network is based on encoder-decoder of U-Net architecture with an encoder of pre-trained VGG16 as an example.}
	\label{fig:barchatSingleClassifiers}
\end{figure}
\subsubsection{Pre-trained CNN Feature Extractors}
For the down-sampling part of the U-Net architecture, different deep CNN-based feature extractors were selected to extract high-level features from the input image. The choice of the feature extractor is based on different modules incorporated into the associated CNN models that successfully achieve the best segmentation performance in the literature. In this study, we selected five Deep CNN architectures as feature extractors, namely ResNet, DenseNet, InceptionV3, InceptionResNetV2 and Squeeze-and-Excitation Networks (SE-ResNeXt), to compare their performance in polyp segmentation task. Residual blocks in ResNet architecture consists of two or three sequential convolutional layers and a supplementary shortcut connection. This shortcut connection adds the output of the previous layer to the output of the next layer, enabling to pass the signal without modification. This architecture helps reduce the degradation of the gradient in deep networks. The inception module creates wider networks rather than deeper by adding filters of three different sizes ( $1\times1$, $3\times3$, and $5\times5$ ) and an additional max-pooling layer. The output is then concatenated together and is sent to the next inception module. Also, before $3\times3$ and $5\times5$ convolutions, an extra $1\times1$ convolution is added to limit the number of input channels. In Dense modules, the previous layer is merged into the future layer by concatenation, instead of using shortcut connections as in ResNet modules. In the Dense module, all feature maps from a layer are connected to all other subsequent layers. SE-ResNeXt introduced an operation that can adaptively recalibrate channel-wise feature responses of each feature map. SE-ResNeXt is the integration of ResNet into squeeze-and-excitation blocks to further improve the accuracy of the network. Inception-ResNet is a hybrid of the Inception architecture with residual connections to boost the representational power of the network. The proposed CNN network based on U-Net architecture with a pre-trained VGG16 feature extractor is illustrated in Fig. 4.

\subsection{Evaluation Criteria}
To measure the performance of the proposed method for polyp segmentation, we employed common segmentation evaluation metrics: Jaccard index, also known as intersection over union (IoU), and Dice similarity score to quantitatively measure similarity and difference between the predicted mask from segmentation model and the ground-truth mask. These metrics are computed by the following:

\begin{equation}
Jaccard\ index\left(A,B\right)=\frac{\left|\ A\ \cap\ B\ \right|}{\left|\ A\ \cup\ B\ \right|}=\ \frac{\left|\ A\ \cap\ B\ \right|}{\left|\ A\ \right|+\left|\ B\ \right|-\left|\ A\ \cap\ B\ \right|\ }
\end{equation}

\begin{equation}
Dice\left(A,B\right)=\ \frac{2\times\left|\ A\ \cap\ B\ \right|}{\left|\ A\ \right|+\left|\ B\ \right|\ }
\end{equation}

Where $A$ represents the output binary mask, produced from the segmentation method and $B$ represents the ground-truth mask, $\cup$ represents union set between $A$ and $B$, and $\cap$ represents the intersection set between $A$ and $B$. 

We also used accuracy to measure the overall accuracy of the segmentation models (binary classification). A high accuracy demonstrates that most of the polyp pixels were classified correctly.

\begin{equation}
Accuracy=\ \frac{TP+TN}{TP+TN+FP+FN}
\end{equation}

True Positive (TP) represents the number of correctly predicted pixels as polyp. False Positive (FP) represents misclassified background pixels as polyp. False Negative (FN) represents misclassified polyp pixels that misclassified as background, and True Negative (TN) represents the background pixels that are correctly classified as background.

\section{Experiments and Results}
\subsection{Experimental Setup}
For this study, we randomly selected 80\% of the CVC-ClinicDB images as the training and validation set and the remaining 20\% for the test set. There is no intersection between the training and test images. To update the weight, we used Adam optimizer with a learning rate, $\beta1$ and $\beta2$ of 10-5, 0.9, 0.999, respectively. The batch size was set to 2, and all models were trained for 50 epochs. Our experiment is implemented in Python using Keras package with Tensorflow as backend and run on Nvidia GeForce GTX 1080 Ti GPU with 11GB RAM. 

\subsection{Results and Discussion}
The accuracy, Dice score and Jaccard index of the obtained results are summarized in Table~\ref{tab:finalResults}. There is a level of variation in the performance of all models. Analyzing Table~\ref{tab:finalResults}, U-Net with DenseNet169 backbone feature extractor outperformed the other approaches, where the U-Net with InceptionResNetV2 backbone feature extractor achieved the second-best results with a slightly lower performance rate. We believe that dense modules, inception modules and also residual blocks as part of U-Net encoder provide an efficient segmentation process and overcome the issue of over-segmentation~\cite{huang2016ensembling}. U-Net with DenseNet169 achieved an accuracy of 99.15\% in comparison to 99.10\% for InceptionResNetV2 architecture. Also, Dice score for DenseNet169 model was 90.87\% compared to 90.42\% for InceptionResNetV2. DenseNet169 also had better results for Jaccard index, 83.82\% compared to 83.16\% for InceptionResNetV2 architecture. 

\begin{table}[ht]
	\centering
	\caption{Evaluation of the segmentation results from different combinations of the pre-trained feature extractors and U-Net architecture.}
	\label{tab:finalResults}
	\begin{tabular}{|l|l|l|l|}
		\hline
		& Accuracy (\%) & DICE (\%) & Jaccard index (\%) \\ \hline
		Baseline U-Net~\cite{ronneberger2015u} & 97.92 & 75.86 & 63.53 \\ 
		SegNet~\cite{badrinarayanan2017segnet} & 95.12 & 68.39 & 61.57 \\ 
		U-Net+ResNet34 & 98.09 & 88.08 & 79.22 \\ 
		U-Net+ResNet50 & 98.77 & 86.06 & 77.62 \\ 
		U-Net+ResNet152 & 98.9 & 87.67 & 79.22 \\ 
		U-Net+DenseNet121 & 98.72 & 85.42 & 77.35 \\
		U-Net+DenseNet169 & \textbf{99.15} &  \textbf{90.87} &  \textbf{83.82} \\ 
		U-Net+DenseNet201 & 98.85 & 87.54 & 80.2 \\ 
		U-Net+InceptionV3 & 99.08 & 89.63 & 81.84 \\ 
		U-Net+InceptionResNetV2 & \underline{99.1} & \underline{90.42} & \underline{83.16} \\ 
		U-Net+SE-ResNeXt50 & 98.79 & 86.61 & 79.05 \\ 
		U-Net+SE-ResNeXt101 & 98.9 & 87.63 & 80.09 \\ \hline
	\end{tabular}
\end{table}

To justify the performance of the ensemble architectures, the performance of baseline U-Net and SegNet architectures are also evaluated and compared with the presented approach. The worst performance is for SegNet with a Jaccard index of 61.57\%, Dice score of 68.39\%, and accuracy of 95.12\%. U-Net with DenseNet169 significantly improves the baseline U-Net up to 15.01\%, and the baseline SegNet architecture up to 22.48\% in terms of Dice score. Moreover, U-Net with DenseNet169 improves baseline U-Net up to 20.29\% and the SegNet architecture up to 22.25\% in terms of Jaccard index. Similar conclusions can be drawn for accuracy metrics. The experimental results indicate the important role of incorporating modules in encoder part of a convolutional segmentation architecture in extracting hierarchical information from input images. 
\begin{table}[ht]
	\centering
	\caption{Comparison of performance of polyp segmentation models on the CVC-ClinicDB dataset.}\label{tab1}
	\label{tab:Results}
	\begin{tabular}{lllll}

		\textbf{Input image} & \textbf{Ground truth} & \textbf{DenseNet169} & \textbf{ResNet50} & \textbf{Baseline U-Net} \\ 
		\includegraphics[scale=0.12]{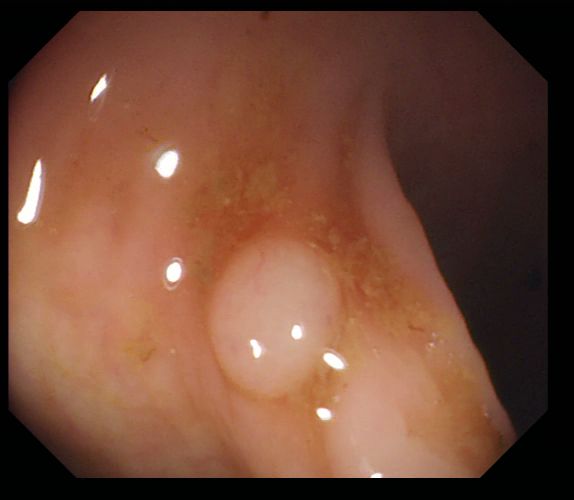} &\includegraphics[scale=0.12]{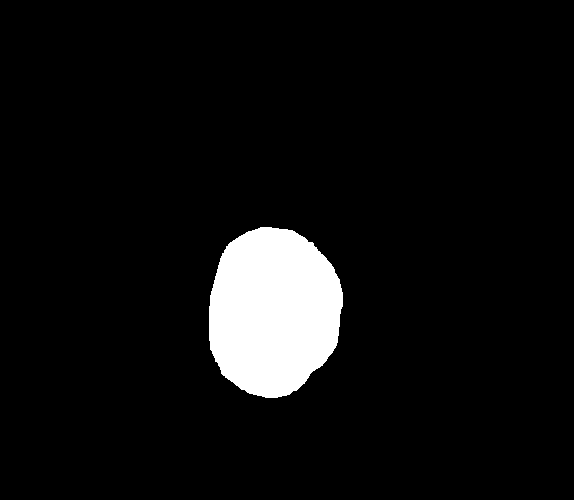} & \includegraphics[scale=0.12]{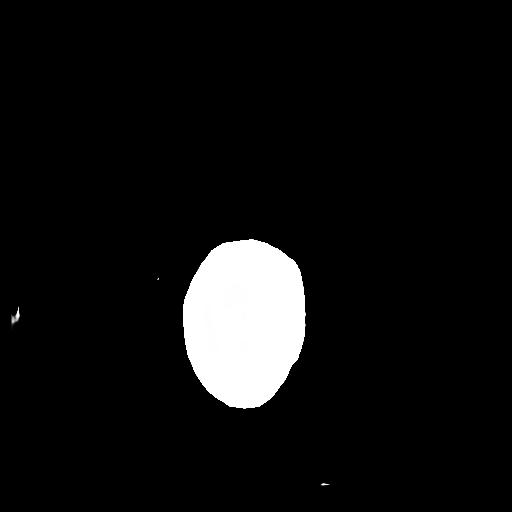} & \includegraphics[scale=0.12]{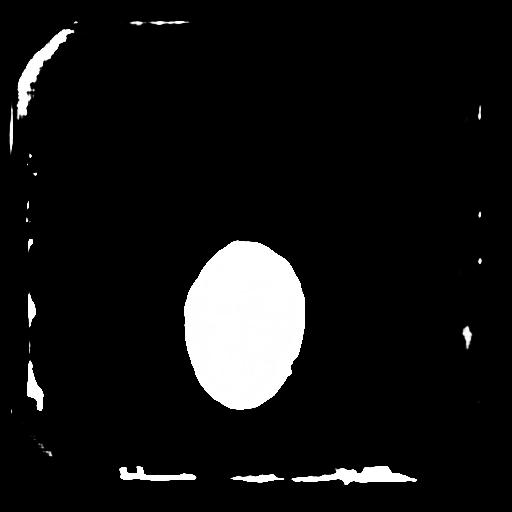} & \includegraphics[scale=0.12]{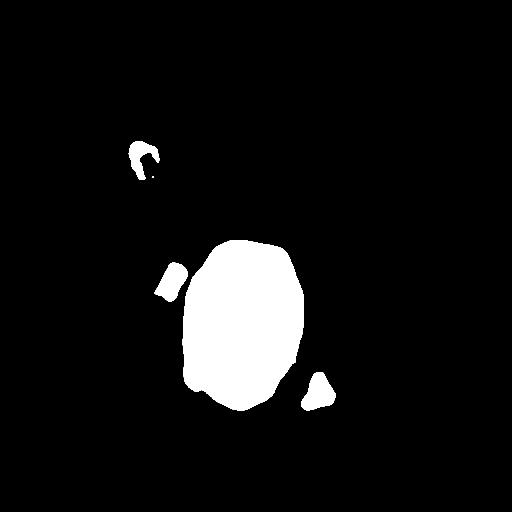} \\ 
		\includegraphics[scale=0.12]{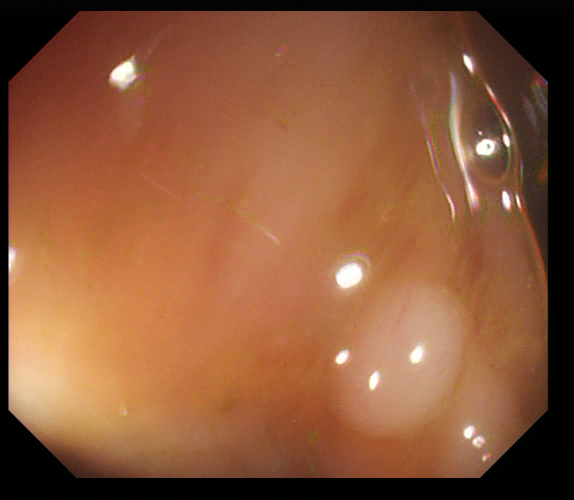} & \includegraphics[scale=0.12]{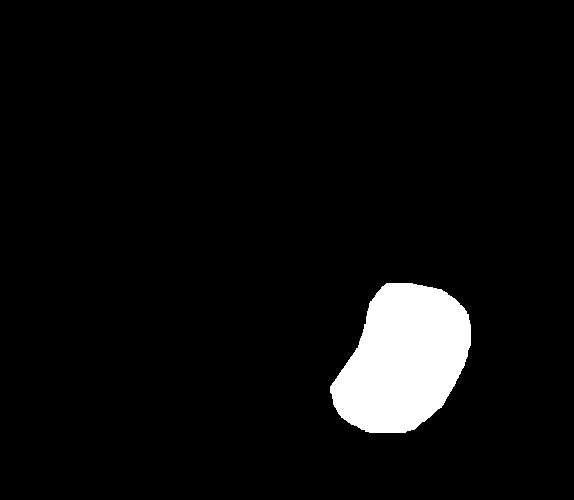} & \includegraphics[scale=0.12]{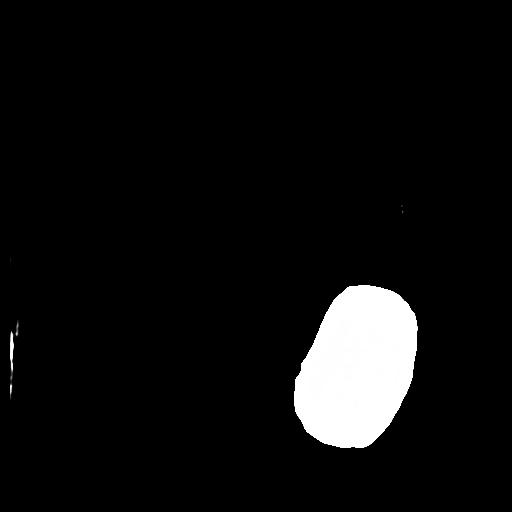} & \includegraphics[scale=0.12]{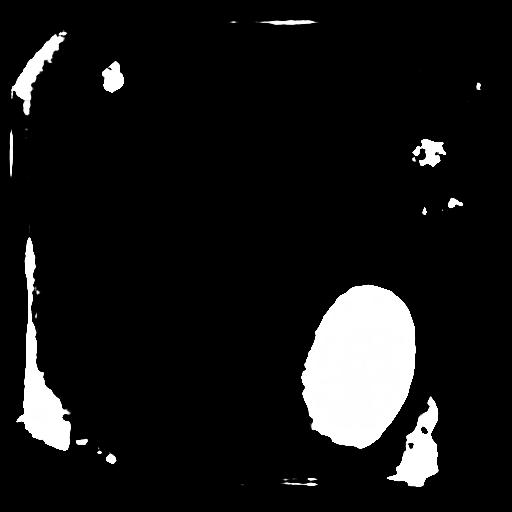} & \includegraphics[scale=0.12]{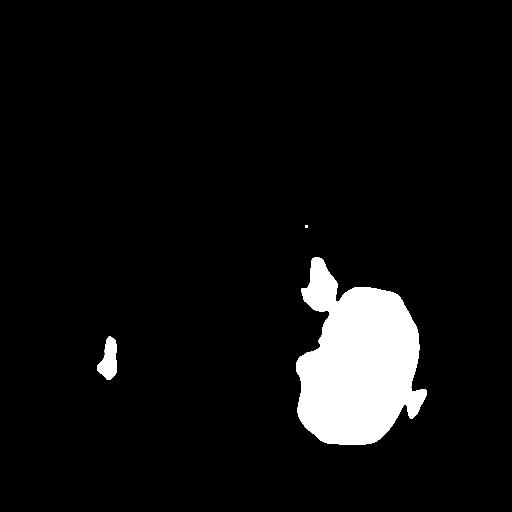} \\ 
		
		\includegraphics[scale=0.12]{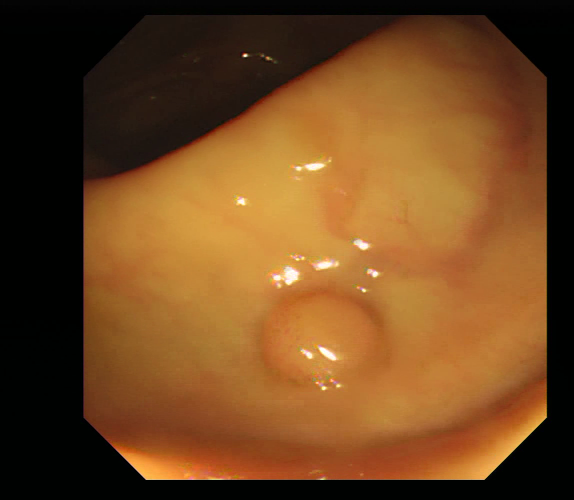} & \includegraphics[scale=0.12]{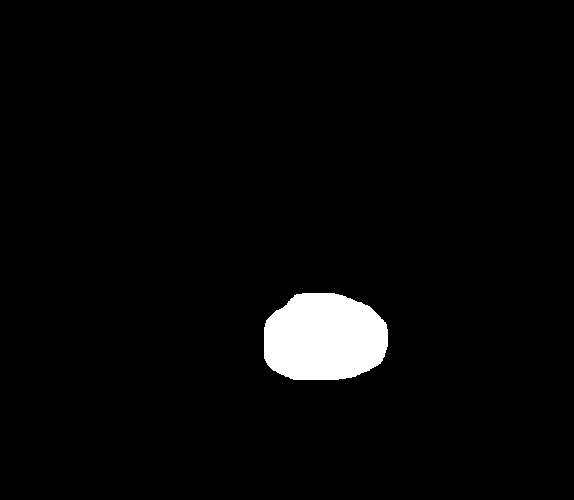} & \includegraphics[scale=0.12]{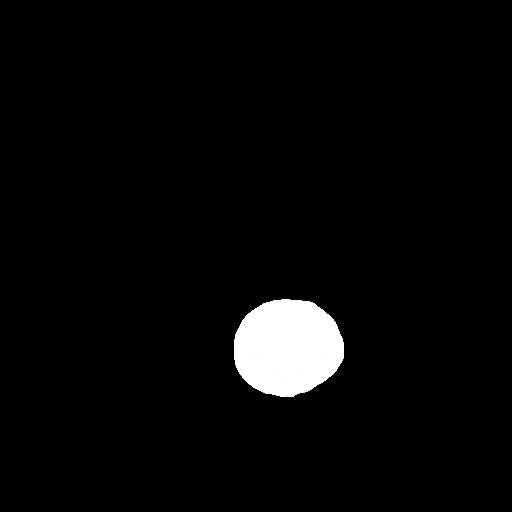} & \includegraphics[scale=0.12]{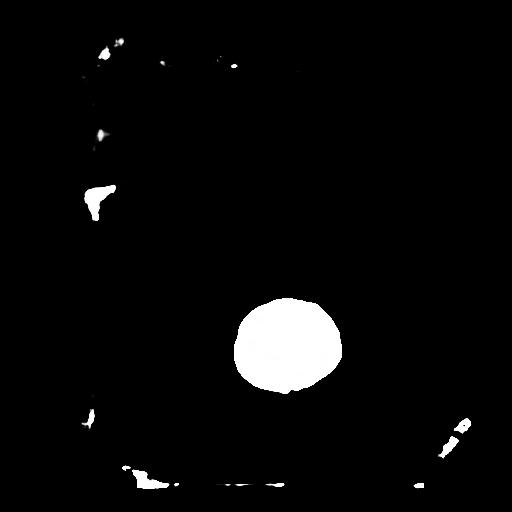} & \includegraphics[scale=0.12]{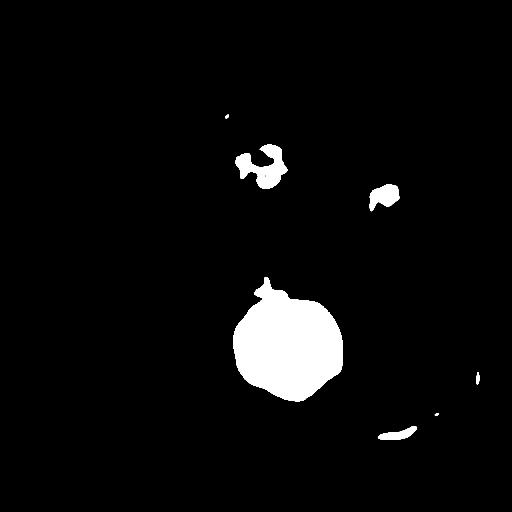} \\
	\end{tabular}
\end{table}

In Table~\ref{tab:Results}, we illustrate three segmentation output results produced by the DenseNet169, ResNet50 and the baseline U-Net model. As the results indicate, the examples selected in the column of DenseNet169 can accurately segment polyps from the background. Also, DensNet169 feature extractor can adequately address different noises present in the input images, including shadows, reflection and blurriness, etc. It should be noted that feature extractors such as ResNet50 and baseline U-Net suffer from over-segmentation. Over-segmentation affects the Dice score and Jaccard index adversely. The main cause of over-segmentation is the low-intensity variations between the foreground and the background and also the lack of enough spatial information. Dense and InceptionResNet modules can eliminate the over-segmentation and effectively segment out polyps with a better performance rate than other models, as demonstrated in Table~\ref{tab:Results}. Table~\ref{tab:comparativeResults} compares the performance of the proposed methods with that of~\cite{li2017colorectal}~\cite{akbari2018polyp}~\cite{qadir2019polyp}~\cite{nguyen2018colorectal}~\cite{kang2019ensemble}. The obtained results were comparable with prior CNN-based methods in the literature, as shown in Table~\ref{tab:comparativeResults}. Both DenseNet169 and InceptionResNetV2 methods show better performance when compared with the existing methods. However, Nguyen and Lee’s approach achieved better results in terms of Jaccard index while the Dice score and accuracy of DenseNet169 and InceptionResNetV2 models outperformed those of Nguyen and Lee’s approach. 

\begin{table}[h]
	\centering
	\caption{Quantitative comparison of the segmentation results with prior CNN-based works on the polyp segmentation task.}
	\label{tab:comparativeResults}
	\begin{tabular}{|l|l|l|l|}
		\hline
		Model & Jaccard index (\%) & Dice score (\%) & Accuracy (\%) \\ \hline
		Li et al.~\cite{li2017colorectal} & - & - & 96.98 \\ 
		Akbari et al.~\cite{akbari2018polyp} & - & 81 & 97.7 \\ 
		Qadir et al.~\cite{qadir2019polyp} & 61.24 & 70.42 &  \\ 
		Nguyen and Lee~\cite{nguyen2018colorectal} & 89.35 & 88.9 & - \\ 
		Kang and Gwak~\cite{kang2019ensemble} & 69.46 & - & - \\ 
		U-Net+DenseNet169 & \textbf{83.82} & \textbf{90.87} & \textbf{99.15} \\ 
		U-Net+InceptionResNetV2 & \underline{83.16} & \underline{90.42} & \underline{99.1} \\ \hline
	\end{tabular}
\end{table}

\section{Conclusion}
In this work, we presented a transfer learning-based encoder-decoder architecture for automated polyp segmentation. The proposed framework consists of a U-Net architecture with different backbone feature extractors, i.e. ResNet, DenseNet, InceptionV3, InceptionResNetV2 and SE-ResNeXt. Our method is validated using a dataset from the CVC-ClinicDB polyp segmentation challenge. The experimental results showed that the proposed ensemble method using DenseNet169 and InceptionResNetV2 feature extractors achieved good results and significantly outperformed the baseline U-Net, and SegNet approaches for polyp segmentation. The main limitation of this work is the limited number of polyp shapes and structures present in the provided dataset, which is a focus of future work. By adding more training samples from external datasets, the deep learning-based segmentation models could gain a better performance and further improve the generalization ability of the network. Our future work will also be dedicated to the investigation of the post-processing methods to reduce the over-segmentation issue.

%

 \bibliographystyle{splncs04}
 \bibliography{references}

\end{document}